\documentstyle{article}[12pt]

\begin{document}

\def\be{\begin{equation}}
\def\ee{\end{equation}}
\def\beqa{\begin{eqnarray}}
\def\eeqa{\end{eqnarray}}
\def\lla{\left\langle}
\def\rra{\right\rangle}
\def\za{\alpha}
\def\zb{\beta}
\def\ssc{\scriptscriptstyle}
\def\bsg{$b \to s \, \gamma\,$}
\def\ga{\mathrel{\raise.3ex\hbox{$>$\kern-.75em\lower1ex\hbox{$\sim$}}}}
\def\la{\mathrel{\raise.3ex\hbox{$<$\kern-.75em\lower1ex\hbox{$\sim$}}}}

\begin{flushright}
NCU-HEP-k031  \\
Oct 2008
\end{flushright}

\vspace*{.5in}

\begin{center}
{\textbf
`AdS$_5$' Geometry Beyond Space-time and 4D Noncommutative Space-time}\\
\vspace*{.5in}
{\textbf  Otto C.W. Kong}\\[.05in]
{\it Department of Physics and Center for Mathematics and
Theoretical Physics, National Central University,  Chung-li, TAIWAN 32054 \\
E-mail: otto@phy.ncu.edu.tw}
\vspace*{1.in}
\end{center}
{\textbf Abstract :}\ \
{We discuss a 4D noncommutative space-time as suggested by the version of quantum (deformed) relativity which provides a classical geometry picture as an `AdS$_5$'. The 4D noncommutative space-time is more like a part of a phase space description, in accordance with the quantum notion -- quantum mechanics talks about only states but not configurations. The `AdS$_5$' picture also illustrates the classical 4D space-time is to be described as part of a bigger geometry beyond space-time at the quantum level. The radically new picture of quantum 'space-time' is expected to provide the basis for a  (still to be formulated) new approach to quantum gravity with fundamental constants (quantum) hbar and Newton's constant G put at a similar level as c, the speed of light.
 }

\vfill
\noindent --------------- \\
$^\star$ Talk presented  at SMP 40 2008 (Jun 25-28), Torun, Poland.


\section{Introduction}
We discuss some aspect of our new approach to think about possible formulation
of the the physics of `quantum space-time' or `quantum gravity'. Our approach
is based on some basic background perspective. We consider there being a true
microscopic/quantum structure of space-time itself and seek a direct description
of that, without going through a scheme of quantization. The latter is in
accordance of various attempts to get to a foundation of quantum physics
with some sort of `hidden variable theory', except that we are ready to 
take the notion much beyond the framework of the usual (classical) geometry
for space-time. It is our belief that new conceptual thinking about what is
`space-time' is necessary to resolve all the issues.
An example of a theory in a similar spirit is offered
by the (Matrix Model) Trace Mechanics published by Adler\cite{A}. It is a new
fundamental dynamics formulated on a matrix model geometry, a typical
class of noncommutative geometry. We do believe the generic noncommutative
geometry, as the natural mathematical generalization of the classical/commutative
geometry \cite{C}, should be the right mathematical setting. In more exact terms,
we believe, or postulate that, 
\begin{center} {\em Non-Commutative Geometry is to Quantum Gravity \\
as Non-Euclidean Geometry is 
to Classical (Einstein) Gravity.} \end{center}
While Adler starts by assuming a particular noncommutative geometry to
be the right space-time background to formulate the fundamental theory
behind quantum mechanics, we want to find a guiding principle on 
how to think about the `space-time' structure itself. A plausible answer
is given by the principle of relativity symmetry stabilization \cite{CO}.

From a pure mathematical point of view, stabilization gives a new symmetry
that has the old one as a contraction limit. The Lorentz symmetry is exactly
a stabilization/deformation of the Galilean counterpart, with the
deformation parameter given by $1/c^2$. Philosophically, one can argue that
only stable symmetries can be scientifically verified to be correct. The
contracted unstable limit is a singular point on the `parameter space',
requiring infinite experimental precision to be confirmed against the
stable symmetry which admits no other parameter value. The pioneering 
work of Snyder in 1947\cite{S} had initiated the idea of a symmetry
deformation being necessary to implement an invariant quantum scale.
The idea has been gaining more attention since the turn of the century
\cite{dr}. 

A `quantum space-time' to be directly described, without going
through any quantization procedure, will have its own relativity. As
suggested by Snyder, simple physics consideration can help identifying
the deformed relativity symmetry. The deformations could be nicely formulated 
as Lie algebra stabilizations \cite{CO}. Following the line of thinking,
we implemented in Ref.\cite{023} a linear realization perspective.  
The linear realization scheme, applied to the Einstein relativity as 
the deformation of Galilean relativity on 3D space, is nothing other than 
the 4D Minkowski space-time picture. Such a mathematically conservative 
approach, however, leads to a very radical physics perspective, that 
space-time is to be described as part of something bigger \cite{023}, 
what we called the quantum world in Ref.\cite{030}. In the latter paper
we identify the quantum relativity symmetry as $SO(2,4)$, with a linear
realization on a 6D classical geometry beyond space-time providing a 
description of a 4D noncommutative (quantum) space-time. The quantum world 
is really the coset space $SO(2,4)/SO(2,3)$, {\it i.e.} the hypersurface 
$\eta_{\ssc \mathcal M\mathcal N} X^{\!\ssc \mathcal M} X^{\!\ssc \mathcal N} = -1$ 
[$\eta_{\ssc \mathcal M\mathcal N} =( 1, -1, -1, -1, -1, 1)$]. For
the lack of a better terminology, we denote it by `AdS$_5$'
\footnote{Note that a version of the table published earlier\cite{030} has 
mistakenly put down the coset space characterized by 
$\eta_{\ssc \mathcal M\mathcal N} X^{\!\ssc \mathcal M} X^{\!\ssc \mathcal N} = -1$ 
here as $SO(2,4)/SO(1,4)$. The careless mistake was propagated to the 
talk presentation file at the SMP 40 conference. The possible inconsistence 
was first brought to the attention of the author by the
group headed by H.Y. Guo at the Institute of Theoretical Physics, CAS, Beijing. 
In relation to that, some physicists may consider the term AdS$_5$ 
inappropriate to be used to named the coset space. We put the term in quotes, 
to alert readers and avoid unnecessary confusion.
}
With the $SO(2,4)$ relativity, fundamental constants  $\hbar$ and $G$ are
essentially putting on exactly the same footing as $c$.

\section{Getting to the relativity symmetry $SO(2,4)$}
In summary, we have the stabilizations and extensions by translations sequence :
{\boldmath \beqa \nonumber
&& ISO(3) \;\; {\rightarrow} \;\; SO(1,3)\;\;\hookrightarrow\;\; ISO(1,3) \\
&& \;\; {\rightarrow} \;\; SO(1,4)
 \;\; \hookrightarrow \;\;ISO(1,4)  \;\; \rightarrow \;\; SO(2,4)\nonumber \eeqa}
The {\boldmath \small $ISO(3)$} algebra is unstable, with 
{\boldmath \small $SO(1,3)$} a possible stabilization result. The only other 
mathematical option is {\boldmath \small $SO(4)$}. It is the physics picture
of the deformation parameter being identified as essentially an upper limit
of admissible speed that fix the right choice. A linear realization adopts
the 4D Minkowski space-time $M^4$ over that of 3D space $I\!\!R^3$ as the 
arena for the relativity theory; {\boldmath \small $SO(1,3)$} is the isometry
of $M^4$. With the linear realization comes the natural extension of the Lorentz
symmetry to Poincar\'e symmetry {\boldmath \small $SO(1,3)$}. The latter is
again an unstable symmetry, to be stabilized to  {\boldmath \small $SO(1,4)$}
with an energy-momentum invariant, the Planck momentum $\kappa\,c$. The physics
behind the stabilization step as well as the last one is summarized in table~1.
In the case of $ct$ in going from $I\!\!R^3$ to $M^4$, a linear realization
asks to have a fifth coordinate $\kappa\,c\,\sigma$ be incorporated to form
the arena for the {\boldmath \small $SO(1,4)$} relativity. The $\sigma$-coordinate
is quite peculiar. It has a space-time geometric signature and a physics dimension
of $(\frac{time}{mass})$. It is neither space or time. After all, whether 
the extra coordinate possibly has the physics meaning of a space-time one
is a question of how one formulate the theory. For example, taking the metric
of the 5D geometry as a gravitational field will be giving $\sigma$ a 
space-time (indeed space rather than time) interpretation. The latter kind
of ready to be formulated theories will be in conflict with the role of
$\sigma$ as seen from the relativity symmetry stabilization physics, or 
equivalently for having the right Einstein limit. The linear realization does
suggest a radical departure of our physics thinking. It provokes even
the question of if we could still formulate dynamics in the way we used to.
Anyway, we have again a natural extension of the symmetry to  
{\boldmath \small $ISO(1,4)$}, and again a stabilization awaits. It may look 
like such a stabilization followed by extension sequence will go on forever.
However, as illustrated in table~1, we take the last stabilization as 
corresponding a `length' invariant, as terminate the sequence. This is like 
implementing a Planck length independent of the Planck mass or momentum.
The idea is that the quantum scale as usually described equally effectively
by Planck mass $\kappa$ and Planck length $\ell$ assumes $\hbar$.
To retrieve $\hbar$ from the symmetry stabilization picture
similar to $c$, one should rather use both Planck momentum and Planck length
for the two deformations following the familiar one with $c$,
getting $\hbar$ from the identity $\hbar= \kappa\,c\,\ell$. 
The $c$ constraint enforces the velocity space to be the coset 
{\boldmath \small $SO(1,3)/SO(3)$}. Likely, the second deformation curves the
momentum space and the last curves the `space-time' arena itself. Hence,
though the quantum relativity symmetry {\boldmath \small $SO(2,4)$} is to
be linearly realized as an isometry of a 6D (beyond space-time) geometry,
the relevant quantum world is only the hypersurface given by 
$\eta_{\ssc \mathcal M\mathcal N} X^{\!\ssc \mathcal M} X^{\!\ssc \mathcal N} = -1$.
Coordinate translations are simply not admissible symmetries. 

\begin{table}[t]
 \caption{The Three Deformations Summarized:
}
\begin{center}
\begin{tabular}{|c|c|c|}    \hline\hline
$\Delta x^i(t) = {v^i} \cdot t$   &   $\Delta x^\mu(\sigma) =
{p^\mu} \cdot \sigma$
 &  $\Delta x^{\ssc A}(\rho) = {z^{\ssc A}} \cdot \rho$  \\
\hline
{$|v^i|\leq c$}           &   {$|p^\mu |\leq \kappa\,c$ }           &    {$| z^{\ssc A} |\leq i\, \ell $} \\
$- \eta_{ij} v^i v^j = c^2 \left(1-\frac{1}{\gamma^2}\right) $
 &   $\eta_{\mu\!\nu} p^\mu p^\nu = \kappa^2 c^2  \left(1-\frac{1}{\Gamma^2}\right)$
      & $\eta_{\ssc \!A\!B} z^{\!\ssc A} z^{\!\ssc B}= - \ell^2 \left(1+\frac{1}{G^2}\right)$     \\
\hline
$M_{{\ssc 0}i}\equiv N_i \sim P_i$                  &   $J_{\mu\ssc 4}\equiv O_\mu \sim P_\mu $
      &   $J_{\ssc \!A\!5}\equiv O'_{\ssc \!A} \sim P_{\ssc \!A} $    \\
{$[N_i, N_j]  \longrightarrow -i\, M_{ij}$}   &   {$[O_{\!\mu}, O_{\!\nu}]  \longrightarrow i\, M_{\mu\nu}$} 
     &  {$[O_{\!\ssc A}^{\prime} , O_{\!\ssc B}^{\prime} ]  \longrightarrow  i\, J_{\ssc \!A\!B}$} \\
\hline $\vec{u}^4=\frac{\gamma}{c}(c , v^i)  $ &
$\vec{\pi}^5=\frac{\Gamma}{\kappa\,c}(p^\mu , \kappa\,c) $  &
        $\vec{X}^6=\frac{G}{\ell}(z^{\ssc A} , \ell) $\\
{$\eta_{\mu\nu} u^\mu u^\nu = 1$}     &   {$\eta_{\ssc A\!B} \pi^{\ssc A} \pi^{\ssc B} = -1 $}
 &   {$\eta_{\ssc \mathcal M\mathcal N} X^{\!\ssc \mathcal M} X^{\!\ssc \mathcal N} = -1$}  \\
{$I\!\!R^3 \!\! \rightarrow SO(1,3)/SO(3)$}       &   {$M^4  \!\!\rightarrow SO(1,4)/SO(1,3)$} &
   {$M^5  \!\!\rightarrow SO(2,4)/SO(2,3)$} 
\\
\hline\hline
\end{tabular}
\end{center}
\end{table}

For the readers who are not familiar with the idea of symmetry stabilization,
we present here a simply argument for an easy appreciation of the 
{\boldmath $ISO(3)$} to {\boldmath $SO(1,3)$} story. Firstly, note that the
only difference in the two algebras is the commutator of two velocity boosts. 
Commutators of Lorentz boost generators are rotation generators, as given by
\[
[N_i, N_j]  = -i\,\frac{1}{c^2}\; M_{ij} \;, 
\]
where $\frac{1}{c^2}$ has been explicitly shown. The commuting algebra of
Galilean boosts is retrieved at the  $\frac{1}{c^2}$ goes to zero limit. The
latter is unstable, as taking any small change in value of the zero 
structural constant changes the algebra. The Lorentz algebra is stable;
different values of $c$ give isomorphic algebra connected by a 
simple scaling. A more direct way is see it is to realize that the mathematics
of symmetry algebra sees no units in physics. The value of $c$ depends, though, 
on our choice of units, we can make it $~3\times 10^{-7}$  (km ps$^{-1}$) or
$~10^{28}$ (A yr$^{-1}$). The symmetry of space-time is of course independent
of what units physicists cooked up to measure things inside. The value of $c$
tells only when we would be in a regime where the Galilean symmetry, as
an approximation of the Lorentz symmetry, is good enough to describe physics.

\section{The new physics picture}
We have the quantum relativity symmetry obtained through the the
Lie algebra stabilization scheme with quite limited physics inputs. To
really construct a theory to be tested experimentally, we need to take
it beyond kinematics. The radical beyond space-time picture posts a
daunting challenge, as we have to figure out the role of the new
$\sigma$ and $\rho$ coordinates in any picture of dynamics. At the most
primitive level, dynamics is a study of motion and motion is characterized
by change of spatial position with respect to time. The 
$\sigma$ and $\rho$ coordinates would have apparently nothing to do 
with motion then. Obviously, understanding the physics role of 
$\sigma$ and $\rho$ will be a key to confront the theoretical challenge
ahead. 

The symmetry deformation scheme does tell us something about 
$\sigma$ and $\rho$. In fact, hidden in the mathematics of table~1
are the sort of definitions for the quantities. As $dt=\frac{dx}{v}$,
we can start thinking about $\sigma$ through  
$p^\mu=\frac{dx^\mu}{d\sigma}$. This is actually another radical 
departure from the conventional mechanics. It is nothing less than 
a new definition of the energy-momentum four-vector. It has been
argued in Ref.\cite{023} that this is admissible so long as the
physics theory guarantees $p^\mu=m\,c\,u^\mu$, the Einstein
energy-momentum at the proper limit. From the latter requirement,
one retrieves for such cases $\sigma=\frac{\tau}{m}$, the proper 
time over the rest mass of an Einstein particle. That is very
interesting for a coordinate with a space-like geometric signature
indeed. Some aspects of the new relativity symmetry transformations
involving $\sigma$, we called momentum boosts, have been discussed
in Ref.\cite{023}. It may characterize some transformations to
quantum frames of reference, with features in basic correspondence
with what has been discussed by some authors of the topic \cite{AK,R}. 
It is interesting to note that parameter essentially the same
as $\sigma$ has been used quite lot in various approaches to
(Einstein) relativistic quantum mechanics in somewhat ambiguous ways.
Our new relativity picture may have to put some of that on solid
theoretical footing, and hence retrieve a better understanding
about the $\sigma$ coordinate \cite{033}. The $\rho$-coordinate
looks simpler. The translational boosts, transformations involving
$\rho$ are like effective translations on the quantum world,
liable to be interpreted as a change in coordinate of the geometric
description. The set of $z^{\ssc A}$'s actually serves as
Beltrami-type coordinate system \cite{030,Guo}. 

Both the $\sigma$ and $\rho$ can also be shown to be closely related to 
the idea of scale transformations. The {\boldmath \small $SO(2,4)$}
is more familiar to theorists to be realized as 4D conformal symmetry.
Checking a possible matching of the symmetry as isometry of the 
6D geometry to that of the conformal symmetry of the 4D space-time 
part gives interesting conclusions \cite{030}. The symmetry matches
to that of the conformal symmetry only on the conformal universe,
a hypersurface given by 
$\eta_{\ssc \mathcal M\mathcal N} X^{\!\ssc \mathcal M} X^{\!\ssc \mathcal N} = 0$.
Translations along the $\sigma$- and $\rho$-coordinate directions 
within the hypersurface are indeed simple scalings. The quantum world
with its characteristic momentum and length scales (quantum scale)
is however not scale invariant. we expect the $\sigma$- and $\rho$-coordinate
translations to maintain a close connection to scale transformations,
if not exactly identical to the latter.

\section{Noncommutative space-time}
The {\boldmath \small $SO(2,4)$} algebra as :
\beqa
&& [M_{\mu\nu}, M_{\lambda\rho}] =
i\hbar\, (\eta_{\nu\!\lambda} M_{\mu\rho} - \eta_{\mu\!\lambda} M_{\nu\rho}
+ \eta_{\mu\rho} M_{\nu\lambda} - \eta_{\nu\rho} M_{\mu\lambda})\;,
\nonumber\\
&& [M_{\mu\nu}, \hat{P}_{\!\lambda}] =  i \hbar \,(\eta_{\nu\!\lambda} \hat{P}_{\!\mu}
    - \eta_{\mu\!\lambda} \hat{P}_{\!\nu}) \;,
\nonumber \\
&& [M_{\mu\nu}, \hat{X}_{\!\lambda}] =  i \hbar \,(\eta_{\nu\!\lambda} \hat{X}_{\!\mu}
    - \eta_{\mu\!\lambda} \hat{X}_{\!\nu}) \;,
\nonumber \\
&& [\hat{X}_{\mu}, \hat{X}_{\!\nu}] =    \frac{i\hbar}{\kappa^2 c^2}  M_{\mu\nu} \;,
\qquad
 [\hat{P}_{\mu}, \hat{P}_{\!\nu}] = -\,\frac{i\hbar}{\ell^2}  M_{\mu\nu} \;,
\nonumber \\
&& [\hat{X}_{\!\mu}, \hat{P}_{\!\nu}] = -i \hbar\, \eta_{\mu\nu} \hat{F} \;,
\qquad
 [\hat{X}_{\!\mu}, \hat{F}] = \frac{-i\hbar}{\kappa^2 c^2} \hat{P}_{\mu} \;,
\qquad [\hat{P}_{\!\mu}, \hat{F}] = \frac{-i \hbar}{\ell^2}
\hat{X}_{\mu}\;,
 \label{tsr1}
\eeqa
($\hbar= {\kappa c \ell}$). This is to be matched to the standard form
\be \label{so} [J_{\ssc \!\mathcal R\mathcal S}, J_{\ssc\! \mathcal
M\mathcal N}] = i\hbar\, ( \eta_{\ssc \mathcal S\mathcal M} J_{\ssc
\mathcal R\mathcal N} - \eta_{\ssc \mathcal R\mathcal M} J_{\ssc
\mathcal S\mathcal N} + \eta_{\ssc \mathcal R\mathcal N} J_{\ssc
\mathcal S\mathcal M} -\eta_{\ssc \mathcal S\mathcal N} J_{\ssc
\mathcal R\mathcal M}) \;, 
\ee
$J_{\ssc \!\mathcal M\mathcal N} = i\hbar\, (x_{\!\ssc \mathcal M}
\partial_{\!\ssc \mathcal N} -x_{\!\ssc  \mathcal N}\, \partial_{\!\ssc  \mathcal M})$.
We identify
\beqa
&& J_{\mu \ssc 4} \equiv - {\kappa\, c} \; \hat{X}_{\mu}
=  i\hbar\, (x_{\mu}\partial_{\ssc 4}
 -x_{\ssc  4}\, \partial_{\mu}) \;,
\nonumber \\
&&  J_{\mu \ssc 5} \equiv -{\ell}\,\hat{P}_{\mu}
=  i\hbar\,  (x_{\mu}\partial_{\ssc 5}
 -x_{\ssc  5}\, \partial_{\mu})\;,
\nonumber \\
&& J_{\ssc 45}  \equiv {\kappa c \ell} \hat{F}=
 i\hbar\,  (x_{\ssc 4}\partial_{\ssc 5} -x_{\ssc  5}\, \partial_{\ssc 4})\;,
\quad J_{\mu\nu} \equiv M_{\mu\nu}\;.
\eeqa
The result gives an interesting interpretation as
suggested by the notation that the generators represent a form of 4D
noncommutative geometry. The sets of $\hat{X}_{\!\mu}$'s and
$\hat{P}_{\!\mu}$'s give indeed natural quantum generalizations of the
classical $x_\mu$'s and $p_\mu$'s (represented as $i\hbar
\partial_\mu$'s here). One can easily check that they do have the 
right classical limit. 

Note that the algebra may also be interpreted as coming from the 
stabilization of the `Poincar\'e + Heisenberg' algebra with $\hat{F}$ 
being the central generator before deformation. On the quantum world, 
$-{\kappa\, c}\,\hat{F}$ is rather just the fifth `momentum' 
component, corresponds to the Beltrami 5-coordinate description. 
It is reasonable to think that both the Poincar\'e and Heisenberg 
algebras must be a part of any `quantum space-time' symmetry. 
Starting with the `Poincar\'e + Heisenberg' algebra then gives 
an alternative clear indication of the need for two step deformations 
from Einstein relativity. Some may complain that the above
interpretation, as the incorporation of the Heisenberg algebra,
is rather a phase space symmetry while we set out to look for
relativity symmetry as linearly realized on some sort of (configuration) 
space(-time). The truth is quantum physics is never about configuration 
space. The latter is next to irrelevant. Quantum mechanics describes 
only evolution of states, without reference to  configurations. Hence, 
the  `quantum space-time' should be more like a quantum phase space 
of the simple space-time. 

We have hence both a 6D classical geometry picture and a 4D
noncommutative geometry picture. This may be somewhat in analog to
the description of a curved geometry  being liable to be described 
within higher a dimensional flat geometry, such as the 3D picture 
of a 2D spherical surface. More investigations into the direction
may help to provide physicists with a more geometric picture
of the generic noncommutative geometry. 

\section{Final remarks}
We have outlined a scheme to think about the physics of
the quantum relativity. We consider the subject a sensible and 
very interesting theoretical endeavor, though conceptually quite
radical in comparison to other approaches in the literature. It
demands creative but careful thinking about physics beyond the 
usual framework. The subject is in a very primitive stage, and the
challenge ahead is formidable. However, the root of our radical
idea of `quantum space-time' as a geometric structure
beyond space-time arise from exactly where Einstein's success
as versus Lorentz failure \cite{M} --- in putting space into beyond space
(space-time). It is amazing to see how much follows from the simple
perspective. We hope to be able to keep making small steps in the
direction forward.

\section{Acknowledgements}
We thank H.-Y. Guo, C.-G. Huang, H.-T. Wu, and their collaborator  
at the Institute of Theoretical Physics (Beijing) for useful discussions.
This work is partially supported by the research Grant No.
96-2112-M-008-007-MY3 from the NSC of Taiwan.

\end{document}